\documentclass{article}
\usepackage[margin=1in]{geometry}
\usepackage{amsmath}
\numberwithin{equation}{section} 
\usepackage{xcolor}

\usepackage{tcolorbox}
\newcommand{\beq}{\begin{equation}}
\newcommand{\eeq}{\end{equation}}
\newcommand{\beqa}{\begin{eqnarray}}
\newcommand{\eeqa}{\end{eqnarray}}
\newcommand{\bdm}{\begin{displaymath}}
\newcommand{\edm}{\end{displaymath}}

\newcommand{\Eq}[1]{Eq.\ (\ref{#1})}
\newcommand{\Eqs}[2]{Eqs.\ (\ref{#1}) and (\ref{#2})}

\newcommand{\Equation}[1]{Equation\ (\ref{#1})}
\newcommand{\Rref}[1]{Ref.\ \cite{#1}}

\newcommand{\Section}[1]{Section\ \ref{#1}}
\newcommand{\Appendix}[1]{Appendix\ \ref{#1}}

\newcommand{\Dp}{\frac{d^3p}{(2\pi)^3}}

\newcommand{\DpEa}{\frac{d^3p}{(2\pi)^3 E_{ap}}}
\newcommand{\partialx}{\partial^{(x)}}
\newcommand{\partialp}{\partial^{(p)}}
\newcommand{\fdista}{f_a}
\newcommand{\fdistas}{f^{(s)}_a}
\newcommand{\fdistazero}{f_{a0}}
\newcommand{\fdistaone}{f_{a1}}
\newcommand{\fdistaonehat}{\hat f_{a1}}

\newcommand{\qa}{q_a}
\newcommand{\Gammaa}{\Gamma_a}
\newcommand{\gammaa}{\gamma_a}
\newcommand{\gammaaone}{\gamma^{(1)}_{a}}

\title{
        Kinetic theory formulation of the $P$- and $CP$-odd terms in the photon
        self-energy in a medium
}

\author{Jos\'e F. Nieves\footnote{nieves@ltp.uprrp.edu},
  John D. Verges\footnote{john.verges@upr.edu}\\[12pt]
  Laboratory of Theoretical Physics, Department of Physics\\
  University of Puerto Rico, R\'{\i}o Piedras, Puerto Rico 00936
}

\date{Jan 2024}

\begin{document}
\maketitle

\begin{abstract}
  In an optically active medium, such as a plasma that contains
  a neutrino background, the left-handed and right-handed polarization
  photon modes acquire different dispersion relations.
  We study the propagation of photons in such a medium, which
  is otherwise isotropic, within the framework of the covariant collissionless
  Boltzmann equation incorporating a term that parametrizes the optical
  activity. Using the linear response approximation, we obtain
  the formulas for the components of the photon polarization tensor,
  expressed in terms of integrals over the
  momentum distribution function of the background particles.
  The main result here is the formula for the $P$- and $CP$-breaking component
  of the photon polarization tensor in terms of the parameter involved in
  the new term we consider in the Boltzmann equation to describe
  the effects of optical activity.
  We discuss the results for some particular cases, such as long-wavelength
  and nonrelativistic limits, for illustrative purposes.
  We also discuss the generalizations of the $P$- and $CP$-breaking
  term we included in the Boltzmann equation. In particular
  we consider the application to a plasma with a neutrino background
  and establish contact with
  calculations of the photon self-energy in those systems
  in the framework of thermal field theory.
\end{abstract}

\section{Introduction and outline}
\label{sec:intro}

It was shown sometime ago, in the context of \emph{thermal field theory} (TFT),
that the general expression for the photon
self-energy in an isotropic medium, consistent with gauge and Lorentz
invariance in four dimensions, may contain a term which signals $P$- and $CP$-
symmetry breaking, either in the Lagrangian, or in the background,
or both\cite{piP,piPCPT}.
The effect of this term is that the dispersion relation of the photon
transverse modes, which would otherwise be degenerate, are split according to
the polarization of the propagating mode.
A consequence of this is the rotation of the
linear polarization of an electromagnetic wave traveling in such media,
or birefringence. These effects arise, for example, when photons propagate
in a medium that contains a neutrino
background\cite{mohanty,nustream2,petropavlova}. 

On the other hand it has also been shown that the effect mentioned can be
described in the context of the classical Maxwell equations in terms of
an additional electromagnetic \emph{constant} $\zeta$, besides the standard
dielectric ($\epsilon$) and magnetic ($\mu$)
constants\cite{thirdc}. A covariant version of this formulation
has also been given\cite{palash1}.
Following \Rref{thirdc} we refer to $\zeta$ as the
\emph{activity constant}, and to this kind of
medium as an \emph{optically active medium}.

Our objective in the present work is to revisit the
study the propagation of photons in such media, in the framework of the
covariant, collisionless, Boltzmann equation, but incorporating a term
that gives rise to the optical activity effects mentioned above.
This approach,  which lies somewhere in the middle between the two approaches
mentioned, that is, TFT on one hand, and a purely
phenomenological description on the other, could be more suitable
than those two approaches in certain situations.

Recently, the optical activity effects on the cosmic microwave
photons as they travel through the medium of the cosmic
neutrino background have been considered in the context
of the cosmic birefringence\cite{cosmicbiref}.
Independently of whether or not the photon-neutrino interactions
are responsible for the observed effects in this context,
or whether another source of the optical activity may be required
as suggested in \Rref{cosmicbiref}, our work may be
useful for further development in this area, and other astrophysical contexts
in which optical activity plays a role.

While in the present work we restrict ourselves to the collisionless
Boltzman equation, modified to incorporate the optical activity effects,
the method we use can be extended to include the effect of
collisions in an optically active plasma, for example by
adapting the techniques used for this purpose in the framework of the Boltzmann
equation for an ordinary plasma (see, e.g., \Rref{rafelski}
and references therein).
For example, in the case of photons propagating in a plasma medium with
a neutrino background, it may be important to include also the effects
of collisional interactions. While such interactions typically lead to
\emph{damping effects}, under the appropriate circumstances can also lead
to growth effects. An example of a growth effect was provided in
\Rref{nustream2}, with regard to the evolution of a magnetic field
perturbation in such a medium. This particular phenomena has been considered
by Semikoz and Sokoloff\cite{semikoz1} as a mechanism for the generation
of large-scale magnetic fields in the early
Universe as a consequence of the neutrino-plasma interactions.
In this sense the present work provides the mechanism for considering the
effects of collisions in an optically active medium on a
firm ground.

In short, our proposal is to consider the relativistic collisionless
Boltzmann equation for a given particle specie ``$a$'', of charge $\qa$,
\beq
\label{kneticeq}
p\cdot\partialx\fdista = -\qa F^{\mu\nu}p_\nu\partialp_\mu\fdista\,,
\eeq
modified by adding a term proportional to the dual electromagnetic tensor
$\tilde F_{\mu\nu}$, as we state precisely below, 
including the explanation of the various symbols that enter in this equation.
The strategy is to use the linear response approach to obtain the expression
for the induced current, and thereby the photon polarization tensor,
in terms of the particle distribution functions,
and study various aspects of the results that could be useful for
applications, such as the corresponding dispersion relations
and the interpretation in terms of the activity constant.
In the present work we restrict ourselves to implement this program for
an isotropic system, that is, the momentum  distribution functions
of all the particle species are isotropic in the rest frame of the system.
The generalization of the approach to other cases, for example
a two-stream component plasma is straightforward
from a conceptual point of view, although of course
the details will be different and they
can be important in specific physical contexts.

The outline of the rest of the paper is as follows. 
In \Section{sec:preliminaries} we present the covariant Boltzmann
equation, that includes the $P$- and $CP$-breaking term
to describe the effects of an optically active medium.
For consistency, we discuss some particular features and consequences
of the equation, including current conservation and the role of the
discrete symmetries of the new term.
In \Section{sec:linearization}, we consider the solution
of the equation, using the standard linearization method.
The expression for the induced current is determined,
in terms of integrals of the momentum distribution functions
of the particles. There we establish contact with the
photon polarization tensor, or equivalently the photon self-energy,
in the TFT language, specifically as used in \Rref{piP},
and the formulas for the components of the polarization tensor
are obtained. The results for the longitudinal and transverse components
of the photon self-energy are the familiar ones. The formula for
the $P$- and $CP$-breaking term is the new result here.
In \Section{sec:discussion}
we discuss some details of the results obtained, and consider
specifically some particular cases (e.g., the long-wavelength
and the nonrelativistic limit) that are useful in many applications
and can serve as benchmark references for more general situations.
We also point out possible generalizations of the $P$-
and $CP$-breaking term we included in the Boltzmann equation, 
in particular how it applies in the context of a plasma 
with a neutrino gas as a background, and establish contact
with calculations of the photon self-energy in such backgrounds
in the framework of TFT\cite{mohanty}.
Possible avenues for extensions and exploration of the
present work are mentioned in \Section{sec:conclusions}.

\section{$P$- and $CP$-breaking kinetic equation}
\label{sec:preliminaries}

Without further preamble, the equation we consider is,
\beq
\label{opaceq0}
p\cdot\partialx\fdista = [-\qa F^{\mu\nu} + \gammaa\tilde F^{\mu\nu}]
p_\nu\partialp_\mu\fdista\,.
\eeq
We will discuss some generalizations of this equation in
\Section{sec:dr:generalization}.
In this expression, $\fdista(x,p)$ is the number density
of the particles specie ``$a$'', of charge $\qa$, in the plasma, expressed
as a function of the four-vectors $x^\mu$ and $p^\mu$, and normalized
as specified below [e.g., see \Eq{na}]. Further, $\tilde F^{\mu\nu}$
is the dual electromagnetic field tensor,
\beq
\label{Fdual}
\tilde F^{\mu\nu} \equiv
\frac{1}{2}\epsilon^{\mu\nu\alpha\beta} F_{\alpha\beta}\,,
\eeq
and we are using the shorthand symbols
\beqa
\partialx_\mu & = & \frac{\partial}{\partial x^\mu}\,,\nonumber\\
\partialp_\mu & = & \frac{\partial}{\partial p^\mu}\,.
\eeqa
We use the conventions such that $g^{\mu\nu}$ has diagonal elements
$(1,-1,-1,-1)$ and $\epsilon^{0123} = +1$.

The parameter $\gammaa$ is a phenomenological parameter,
which in this approach is unknown. However \Eq{opaceq0},
or its generalizations, as we will discuss, parametrize
effectively the results of the calculations of optical activity in some
systems, such as those considered in the references already cited
(e.g., Refs, \cite{mohanty,nustream2,petropavlova}). In this sense
this approach serves as a bridge between those calculations on one hand,
and a pure phenomenological description in terms of the electrodynamics
equations on the other (e.g., Refs. \cite{piP,thirdc,palash1}).

Before entering in the practical calculations, there are various
aspects of these equations that are worth discussing. We consider them
below.

\subsection{Current conservation}
\label{sec:conservation}

The electromagnetic current density is
\beq
\label{j}
j_\mu = \sum_a \qa J_{a\mu}\,,
\eeq
where
\beq
\label{Jcov}
J_{a\mu} = 2\int\frac{d^4p}{(2\pi)^3} \delta(p^2 - m^2_a)\theta(p\cdot u)
p_\mu\fdista\,.
\eeq
We have introduced the velocity four-vector of the medium, $u^\mu$,
which has components
\beq
\label{u}
u^\mu = (1,\vec 0)\,,
\eeq
in the medium's own rest frame. Notice that
\beq
\delta(p^2 - m^2_a)\theta(p\cdot u) =
\frac{1}{2E_{ap}}\delta(p^0 - E_{ap})\,,
\eeq
with
\beq
E_{ap} = \sqrt{|\vec p|^2 + m^2_a}\,,
\eeq
since $\theta(p\cdot u) = 0$ for the negative solution $p^0 = -E_{ap}$.
Therefore \Eq{Jcov} reduces to
\beq
\label{Jstandard}
J_{a\mu} = \int\frac{d^3p}{(2\pi)^3 E_{ap}}p_\mu \fdista\,,
\eeq
which is the conventional expression for the current density
four-vector of each specie. In particular, its zeroth
component is the particle number density of each specie,
\beq
\label{na}
n_a = J^0_{a} = \int\frac{d^3p}{(2\pi)^3}\fdista\,.
\eeq

We now consider the divergence of the current density. From \Eq{Jcov},
\beq
\partialx\cdot J_a =
2\int\frac{d^4p}{(2\pi)^3} \delta(p^2 - m^2_a)\theta(p\cdot u)
(p\cdot\partialx\fdista)\,,
\eeq
and using \Eq{opaceq0}
\beq
\label{divJa}
\partialx\cdot J_a = A^{\mu\nu}_a J_{a\mu\nu}\,,
\eeq
where we have defined
\beq
\label{Amunu}
A^{\mu\nu}_a = -q_aF^{\mu\nu} + \gammaa \tilde F^{\mu\nu}\,,
\eeq
and
\beqa
\label{Jmunu}
J_{a\mu\nu} & \equiv & \int d^4p \, \delta(p^2 - m^2_a)\theta(p\cdot u)
p_\nu\partialp_\mu\fdista\nonumber\\
& = & -\int d^4p\,\fdista\, \partialp_\mu
\left[p_\nu \delta(p^2 - m^2_a)\theta(p\cdot u)\right]\,.
\eeqa
In the second equality in \Eq{Jmunu} we have integrated by parts,
dropping the surface term since $\fdista$ vanishes for infinite momentum.
Current conservation is a consequence of the fact that, while $A^{\mu\nu}_a$
is antisymmetric,
\beq
A^{\mu\nu}_a = - A^{\nu\mu}_a\,,
\eeq
$J_{a\mu\nu}$ is, as we show below, symmetric,
\beq
\label{Jsymmetric}
J_{a\mu\nu} = J_{a\nu\mu}\,.
\eeq
Therefore,
\beq
\label{AmunuJmunu}
A^{\mu\nu}_a J_{a\mu\nu}  = 0\,,
\eeq
which together with \Eq{divJa} imply that
\beq
\label{continuity}
\partialx\cdot J_a = 0\,.
\eeq

The proof of \Eq{Jsymmetric} is by straightforward algebra.
Taking derivative that appears in the integrand of \Eq{Jmunu} we obtain,
\beqa
\label{derivativeinJmunu}
\partialp_\mu\left[p_\nu \delta(p^2 - m^2_a)\theta(p\cdot u)\right] & = &
g_{\mu\nu} \delta(p^2 - m^2_a)\theta(p\cdot u)\nonumber\\
&&\mbox{} +
p_\nu \partialp_\mu
\left[\delta(p^2 - m^2_a)\theta(p\cdot u)\right]\,,
\eeqa
and for the second term in \Eq{derivativeinJmunu}
\beq
\partialp_\mu\left[\delta(p^2 - m^2_a)\theta(p\cdot u)\right] =
p_\mu\delta^\prime(p^2 - m^2_a)\theta(p\cdot u) +
u_\mu\delta(p^2 - m^2_a)\delta(p\cdot u)\,.
\eeq
In the last formula, the product of the two \emph{delta} functions
give zero because the two conditions,
\beq
p^{0} = \pm\sqrt{|\vec p|^2 + m^2_a}\,,
\eeq
and
\beq
p^0 u^0 = \vec p\cdot\vec u\,,
\eeq
cannot be satisfied simultaneously. Therefore we get
\beq
J_{a\mu\nu} = -\int d^4p\,f\,\theta(p\cdot u)\left[
g_{\mu\nu} \delta(p^2 - m^2_a) + p_\mu p_\nu\delta^\prime(p^2 - m^2_a)\right]\,,
\eeq
which explicitly verifies \Eq{Jsymmetric}.

\subsection{Modification of the Boltzmann equation}

If the distribution function is treated as a function only of $\vec p$
rather than $\vec p$ and $p^0$ separately, in other words we explicitly set
\beq
\fdistas(t,\vec x, \vec p) \equiv \left.\fdista(p)\right|_{p^0 = E_p}\,,
\eeq
it is well known that the covariant equation given
in \Eq{kneticeq} is equivalent to (see, e.g., \Rref{melrose})
\beq
\label{stdnoncovariant}
\partial_t\fdistas + \vec v_a\cdot\nabla_x\fdistas =
-\vec F_a\cdot\nabla_p\fdistas\,,
\eeq
with
\beq
\vec F_a = q_a(\vec E + \vec v_a\times\vec B)\,.
\eeq
\Equation{stdnoncovariant} is the standard form of the Boltzmann
equation for charged particles in an electromagnetic field.

In the case that $\fdista$ satisfies \Eq{opaceq0}
the corresponding equation, analogous to \Eq{stdnoncovariant},
is
\beq
\label{opacnoncovariant}
\partial_t\fdistas + \vec v_a\cdot\nabla_x\fdistas = (-\vec F_a + \vec G_a)
\cdot\nabla_p\fdistas\,,
\eeq
where $\vec F_a$ is given above, and
\beq
\vec G_a = \gammaa(\vec B - \vec v_a\times\vec E)\,.
\eeq
The result given in \Eq{opacnoncovariant}
follows easily from the fact that the elements of $\tilde F_{\mu\nu}$
are obtained from $F_{\mu\nu}$ by making the replacement
\beqa
\vec E & \rightarrow & \vec B\,,\nonumber\\
\vec B & \rightarrow & -\vec E\,,
\eeqa
which implies that $\vec G_a$ is obtained from $\vec F_a$ by making
the same replacement.
Thus, for example, in the presence of only a magnetic field $\vec B$,
the equation is
\beq
\partial_t\fdistas + \vec v_a\cdot\nabla_x\fdistas =
(-q_a\vec v_a\times\vec B + \gammaa\vec B) \cdot\nabla_p\fdistas\,.
\eeq

A quick observation that follows from \Eq{opacnoncovariant} is that
some discrete space-time symmetries are broken in the system
when the $\tilde F_{\mu\nu}$ term is present in the Boltzmann equation.
This is obvious, for example for parity ($P$), from the fact
that $\vec E$ and $\vec B$ have opposite phase under a $P$
transformation, and therefore the same  holds for $\vec F$ and $\vec G$.

\section{Linearization of the kinetic equation and $\pi_{\mu\nu}$}
\label{sec:linearization}

\subsection{Linearization and the induced current}

The dispersion relations for the propagating photons are obtained
by linearizing the kinetic equation. We put
\beq
\label{flinear}
\fdista = \fdistazero + \fdistaone + \cdots\,,
\eeq
where $\fdistazero$ is the equilibrium distribution, and $\fdistaone$ is linear
in $F^{\mu\nu}$. Substituting \Eq{flinear} in \Eq{opaceq0}, 
and retaining only terms that are linear in $F^{\mu\nu}$, gives
\beq
\label{opaceq0linear}
p\cdot\partialx\fdistaone =
[-\qa F^{\mu\nu} + \gammaa \tilde F^{\mu\nu}] p_\nu\partialp_\mu\fdistazero\,.
\eeq
The next step is to consider the momentum space equation
corresponding to \Eq{opaceq0linear}. Denoting the
wave vector by $k^\mu$, the momentum space
equation is obtained from \Eq{opaceq0linear} by making the replacements
\beqa
\label{kreplacements}
F_{\mu\nu} & \rightarrow & f_{\mu\nu}\,,\nonumber\\
\tilde F_{\mu\nu} & \rightarrow & \tilde f_{\mu\nu}\,,\nonumber\\
\fdistaone & \rightarrow & \fdistaonehat\,,\nonumber\\
\partialx_\mu\fdistaone & \rightarrow & -ik_\mu \fdistaonehat\,,
\eeqa
with the understanding that the functions on the right-hand side
are the Fourier transforms of those on the left. In particular,
\beq
\label{fdual}
\tilde f^{\mu\nu} = \frac{1}{2}\epsilon^{\mu\nu\alpha\beta} f_{\alpha\beta}\,,
\eeq
in correspondence with \Eq{Fdual}. Furthermore, remembering that
we are considering an isotropic system, $\fdistazero$ is a function
only of
\beq
{\cal E} \equiv p\cdot u\,,
\eeq
in which case
\beq
\label{isotropiccondition}
\partialp_\mu\fdistazero = u_\mu \fdistazero^\prime\,,
\eeq
where
\beq
\label{f0prime}
\fdistazero^\prime \equiv \frac{\partial\fdistazero}{\partial{\cal E}}\,.
\eeq
With these substitutions, the momentum space equation corresponding
to \Eq{opaceq0linear} is,
\beq
\label{opaceq0lineark}
(-ik\cdot p)\fdistaonehat = \left[-q_a f^{\mu\nu} +
\gammaa \tilde f^{\mu\nu}\right]u_\mu p_\nu\fdistazero^\prime\,,
\eeq
which gives
\beq
\label{opaceq0linearksol}
\fdistaonehat = \left(\frac{i}{k\cdot p}\right)\left[
-q_a f^{\mu\nu} + \gamma_a \tilde f^{\mu\nu}\right]
u_\mu p_\nu\fdistazero^\prime\,.
\eeq

The induced current is obtained from \Eqs{j}{Jstandard} using
\Eqs{flinear}{opaceq0linearksol}.
The terms containing the equilibrium distributions $\fdistazero$ do
not contribute. This is most easily seen by going to the medium's
own rest frame. Due to the isotropy condition, the vector current
density $\vec j_a$ is zero, while the total $j^0$ is zero assuming
that the charge density is zero in equilibrium. Thus,
\beq
\label{jlinearf}
j_\mu = i\sum_a\qa\left[-\qa f^{\lambda\nu} +
  \gammaa\tilde f^{\lambda\nu}\right]u_\lambda I^{(1)}_{a\mu\nu}\,,
\eeq
where
\beq
\label{I1munu}
I^{(1)}_{a\mu\nu} = \int\DpEa\frac{\fdistazero^\prime}{k\cdot p}p_\mu p_\nu\,.
\eeq

The next step is to write \Eq{jlinearf} in terms of the vector potential
rather than the field. We consider separately the two terms in \Eq{jlinearf}.

\subsubsection{$f^{\mu\nu}$ term}

For the term with $f^{\mu\nu}$, using
\beq
\label{fmunuA}
f_{\alpha\beta} = -i[k_\alpha A_\beta - k_\beta A_\alpha]\,,
\eeq
we have
\beq
u^\alpha p^\beta f_{\alpha\beta} =
-i[(k\cdot u)p_\nu - (k\cdot p)u_\nu]A^\nu\,,
\eeq
which gives
\beq
\label{jfterm}
f^{\lambda\nu} u_\lambda I^{(1)}_{a\mu\nu} = -i[(k\cdot u)I^{(1)}_{a\mu\nu} -
  I^{(2)}_a u_\mu u_\nu]A^\nu\,,
\eeq
where $I^{(1)}_{a\mu\nu}$ is given in \Eq{I1munu} and
\beq
\label{I2}
I^{(2)}_{a} = \int\DpEa (p\cdot u)\fdistazero^\prime\,.
\eeq
To arrive at \Eq{jfterm} we have used again the fact that we are considering
an isotropic system so $\fdistazero$ is a function of $p\cdot u$,
and therefore
\beq
\label{I2muI2}
\int\DpEa\fdistazero^\prime p_\mu = I^{(2)}_{a} u_\mu\,.
\eeq
For the sake of completeness, we mention that the contribution to the current
from the term given in \Eq{jfterm} is transverse by itself, as can be easily
verified explicitly by multiplying by $k^\mu$ and using \Eq{I2muI2}.

\subsubsection{$\tilde f^{\mu\nu}$ term}

We now derive the relation analogous to \Eq{jfterm} for the $\tilde f^{\mu\nu}$
term. From \Eqs{fdual}{fmunuA}, relabeling some of the Lorentz indices,
it follows that
\beq
\label{jftildeterm1}
\tilde f^{\lambda\nu} u_\lambda I^{(1)}_{a\mu\nu} =
-i\epsilon^{\lambda\nu\alpha\beta}k_\alpha u_\beta
A_\nu I^{(1)}_{a\mu\lambda}\,,
\eeq
where $I^{(1)}_{a\mu\nu}$ is defined in \Eq{I1munu}.
Since $I^{(1)}_{a\mu\nu}$ is a symmetric in $\mu,\nu$, and depends only
on $k^\mu$ and $u^\mu$, it is of the form
\beq
I^{(1)}_{a\mu\nu} = C g_{\mu\nu} + X_1 u_\mu u_\nu + X_2 k_\mu k_\nu +
X_3(k_\mu u_\nu + u_\mu k_\nu)\,.
\eeq
Only the $C$ term contributes in \Eq{jftildeterm1} and therefore
\beq
\label{jftildeterm2}
\tilde f^{\lambda\nu} u_\lambda I^{(1)}_{a\mu\nu} =
-iC_a\epsilon_{\mu\nu\alpha\beta}k^\alpha u^\beta A^\nu\,.
\eeq
$C_a$ can in turn be written in terms of the
tensor $R^{\lambda\rho}$ defined in \Eq{RQP} as
\beq
\label{C}
C_a = \frac{1}{2}R^{\lambda\rho}I^{(1)}_{a\lambda\rho}\,.
\eeq

\subsection{$\pi_{\mu\nu}$ and $\pi_{T,L,P}$}

Using \Eqs{jfterm}{jftildeterm2} in \Eq{jlinearf}, the induced current is
then expressed in the form $j_\mu = -\pi_{\mu\nu}A^\nu$ [\Eq{jpi}], with
\beq
\label{piexplicit}
\pi_{\mu\nu} = \sum_a\left\{\qa^2[(k\cdot u)I^{(1)}_{a\mu\nu} -
  I^{(2)}_a u_\mu u_\nu] - \qa\gammaa C_a
  \epsilon_{\mu\nu\alpha\beta}k^\alpha u^\beta\right\}\,.
\eeq
This expression for $\pi_{\mu\nu}$ can be decomposed
in terms of $\pi_{T,L,P}$ as given in \Eq{pimunuparam}.
$\pi_P$ can be written by inspection, while $\pi_{T,L}$ can be
obtained by projecting the term in square brackets in \Eq{piexplicit},
which is symmetric and transverse, with $R^{\mu\nu}$ and $Q^{\mu\nu}$,
using \Eqs{RQPproducts}{RQPnormalization}. Thus,
\beqa
\label{piTLP}
\pi_L & = & -\sum_a\qa^2\left(\frac{k^2 D_a}{\kappa^2}\right)\,,\nonumber\\
\pi_T & = & \sum_a \qa^2 (k\cdot u) C_a
\,,\nonumber\\
\pi_P & = & \sum_a i\qa\gammaa\kappa C_a\,,
\eeqa
where $C_a$ has been defined in \Eq{C}, while
\beqa
\label{D}
D_a & = & u^\mu u^\nu[(k\cdot u)I^{(1)}_{a\mu\nu} - I^{(2)}_a u_\mu u_\nu]\,.
\eeqa
Using \Eqs{I1munu}{I2}, the integral formulas for $C_a$ and $D_a$
can be written in the form
\beqa
\label{CD}
C_a & = & \frac{1}{2}\int\DpEa\left\{
p^2 - (p\cdot u)^2 +
\frac{1}{\kappa^2}[p\cdot k - (k\cdot u)(p\cdot u)]^2\right\}
\frac{\fdistazero^\prime}{k\cdot p}\,,\nonumber\\
D_a & = & \int\DpEa\{(k\cdot u)(p\cdot u)^2 - (p\cdot u)(k\cdot p)\}
\frac{\fdistazero^\prime}{k\cdot p}\,.
\eeqa

We consider the evaluation of the integrals for $C_a$ and $D_a$
in the medium's rest frame, and thus we set
\beq
u^\mu = (1,\vec 0)\,.
\eeq
We decompose $k^\mu$ in the form
\beq
\label{krestframe}
k^\mu = (\omega,\vec\kappa)\,,
\eeq
and in the integrands we will write
\beq
p^\mu = (E_{ap},\vec p)\,,
\eeq
with the understanding that we are working in the rest frame of the medium.
In particular $\fdistazero$ is a function only of $E_{ap}$, and
\beq
\fdistazero^\prime = \frac{\partial \fdistazero}{\partial E_{ap}}\,.
\eeq
Thus, for example, the numerator of the integrand for $D_a$ in \Eq{CD}
\beq
(k\cdot u)(p\cdot u)^2 - (p\cdot u)(k\cdot p) =
\omega E^2_{ap} - E_{ap}(\omega E_{ap} - \vec\kappa\cdot\vec p) =
E_{ap}\vec\kappa\cdot\vec p\,,
\eeq
while the denominator can be written as
\beq
E_{ap}(\omega - \vec\kappa\cdot\vec v_a)\,,
\eeq
with
\beq
\vec v_a = \frac{\vec p}{E_{ap}}\,.
\eeq
Substituting these in the expression for $D_a$ in \Eq{CD},
\beq
D_a = \int\Dp\left[
  \frac{\vec\kappa\cdot\vec v_a}{\omega - \vec\kappa\cdot\vec v_a}
  \right]\fdistazero^\prime\,,
\eeq
and similarly,
\beq
C_a = -\frac{1}{2}\int\Dp \frac{\vec v^{\,2}_{a\perp}}
{\omega - \vec\kappa\cdot\vec v_a}\fdistazero^\prime\,,
\eeq
where
\beq
\vec v_{a\perp} =
\vec v_a - \frac{1}{\kappa^2}(\vec\kappa\cdot\vec v_a)\vec\kappa\,.
\eeq
Although we do not indicate it explicitly, it is understood that
in these formulas for $C_a$ and $D_a$, as well as in the
formulas for $\pi_{L,T,P}$ given below, the singularity of the
integrand at $\omega = \vec\kappa\cdot\vec v_a$ is to be handled
by making the replacement
\beq
\omega \rightarrow \omega + i0^{+}\,,
\eeq
as usual.

From \Eq{piTLP} we then obtain
\beqa
\pi_L & = & -\frac{k^2}{\kappa^2}\sum_a \qa^2\int\Dp\left[
  \frac{\vec\kappa\cdot\vec v_a}{\omega - \vec\kappa\cdot\vec v_a}
  \right]\fdistazero^\prime\,,\nonumber\\
\pi_T & = & -\frac{\omega}{2}\sum_a q^2_a
\int\Dp \frac{\vec v^{\,2}_{a\perp}}
{\omega - \vec\kappa\cdot\vec v_a}\fdistazero^\prime\,,\nonumber\\
\pi_P & = & -\frac{i\kappa}{2}\sum_a q_a\gammaa
\int\Dp \frac{\vec v^{\,2}_{a\perp}}
{\omega - \vec\kappa\cdot\vec v_a}\fdistazero^\prime\,.
\eeqa
The corresponding expressions for $\epsilon_{\ell,t}$ obtained from
\Eq{piepsilonrelations} reproduce the standard classic results, e.g.,
\beq
\epsilon_\ell - 1 = \frac{1}{\kappa^2}\sum_a \qa^2\int\Dp\left[
  \frac{\vec\kappa\cdot\vec v_a}{\omega - \vec\kappa\cdot\vec v_a}
  \right]\fdistazero^\prime\,,
\eeq
and similarly for $\epsilon_t$.
%
%
On the other hand, the formula for $\pi_P$, and the corresponding
formula for $\epsilon_p$ obtained from \Eq{piepsilonrelations}, are new.

\section{Discussion}
\label{sec:discussion}

\subsection{Dispersion relations in the long-wavelength limit}

The longitudinal dispersion relation is not affected by the $\gamma_a$
terms, as already indicated. Therefore, we focus on the transverse ones
(i.e., polarizations perpendicular to $\vec\kappa$) which involve
both $\pi_T$ and $\pi_P$. Moreover, we consider specifically the
long-wavelength limit,
\beq
\omega \gg \kappa v_a\,,
\eeq
which is a particularly useful and representative of more general
situations.

Using the fact that we are considering the case that
the distribution functions $\fdistazero$ are isotropic,
by straightforward manipulation of the integrand,
\beqa
C_a(\omega,\kappa\rightarrow 0) & = & \frac{\Omega^2_a}{\omega}\,,
\eeqa
where
\beq
\Omega^2_a = \int\DpEa\left(1 - \frac{v^2_a}{3}\right)\fdistazero\,.
\eeq
For reference, recall that the plasma frequency ($\omega_{pl,a})$
of each specie is given by
\beq
\label{defplasmafreq}
\omega^2_{pl,a} = q^2_a\Omega^2_a\,.
\eeq
For example, in the nonrelativistic limit,
\beq
\label{defplasmafreqNR}
\Omega^2_a = \frac{n_{a0}}{m_a}\,,
\eeq
where $n_{a0}$ is the equilibrium particle number density of the
specie [see \Eq{na}]. From \Eq{piTLP} we then obtain in the
long-wavelength limit
\beq
\label{piPlw}
\pi_P(\omega,\kappa\rightarrow 0) = \frac{i\kappa\gamma_P}{\omega}\,,
\eeq
and the well-known result
\beqa
\label{piLTlw}
\pi_T(\omega,\kappa\rightarrow 0) & = & \Omega^2_0\,,
\eeqa
where
\beq
\Omega^2_0 = \sum_a q^2_a \Omega^2_a\,,
\eeq
and
\beq
\label{gammaP}
\gamma_P = \sum_a q_a \gamma_a\Omega^2_a\,.
\eeq

As reviewed in \Appendix{sec:appendixA}, the transverse dispersion relations
are determined as the solutions of
\beq
k^2 - (\pi_T + \lambda\pi_P) = 0\,,\qquad (\lambda = \pm)\,.
\eeq
Substituting in this equation the results for $\pi_P$ and $\pi_T$
given in \Eqs{piPlw}{piLTlw}, respectively,
in the long-wavelength limit the equation becomes
\beq
\label{dreq2}
\omega^2 - (\kappa^2 + \Omega^2_0) - \frac{i\lambda\kappa\gamma_P}{\omega}
= 0\,.
\eeq
In the limit $\gamma_P \rightarrow 0$ we obtain the standard
transverse solutions,
\beq
\omega = \omega_T(\kappa) \equiv \sqrt{\kappa^2 + \Omega^2_0}\,,
\eeq
for either polarization. In the more general situation, assuming
\beq
|\gamma_P| \ll \frac{2\omega^3_T(\kappa)}{\kappa}\,, 
\eeq
the solutions are
\beq
\omega(\kappa) = \omega_T(\kappa) +
i\frac{\lambda\kappa\gamma_P}{2\omega^2_T(\kappa)}\,.
\eeq
The dispersion relations are such that one polarization mode is damped
(absorption by the medium) while the other one grows (emission
by the medium).
Which is one or the other depends on the sign of $\gamma_P$, which
in turn depends on the relative signs and values of the $q_a\gammaa$
terms in \Eq{gammaP}. 

\subsection{Generalization}
\label{sec:dr:generalization}

\Equation{opaceq0} is probably the simplest equation of the kind
we are discussing, but there are some possible generalizations.
Here we mention some of them. We will write them in the generic form
\beq
\label{opacgeneric}
p\cdot\partialx\fdista = [-\qa F^{\mu\nu} + \Gammaa^{\mu\nu}]
p_\nu\partialp_\mu\fdista\,.
\eeq
The requirement is that $\Gamma^{\mu\nu}$ must contain, in some form,
the dual tensor $\tilde F^{\alpha\beta}$, and that it is antisymmetric
in $\mu,\nu$ so that the proof of current conservation given in
\Section{sec:conservation} applies in this case as well,
with the identification of $A^{\mu\nu} = -\qa F^{\mu\nu} + \Gamma^{\mu\nu}$
in place of \Eq{Amunu}.

One possibility is
\beq
\label{opaceq1}
\Gammaa^{\mu\nu} = \gammaaone(u\cdot\partialx)\tilde F^{\mu\nu}\,,
\eeq
where $\gammaaone$ is a constant parameter. The steps
to arrive at \Eq{opaceq0linearksol} for $\fdistaonehat$
apply also in this case, with the identification
\beq
\label{correponddencecase1}
\gammaa = -i(k\cdot u)\gammaaone\,.
\eeq
Thus for example, from \Eqs{piPlw}{gammaP}, we can see that in the
long wavelength limit this gives a contribution to $\pi_P$ of the form
\beq
\kappa q_a\gammaaone\Omega^2_a\,.
\eeq
The main qualitative difference, relative to the case in which
$\gammaa$ is independent of $\omega$, is that in the present case
the corresponding contribution to $\pi_P$ is real, which in turn
produces a real term in the dispersion relations of opposite
sign for the $(\pm)$ polarizations.

More generally, we can consider the equation
\beq
\label{Gammamunukernel}
\Gammaa^{\mu\nu} =
\int\,d^4x^\prime\Gammaa(x - x^\prime)\tilde F^{\mu\nu}(x^\prime)\,.
\eeq
The momentum space equation in the linear approximation for $\fdistaonehat$
is again \Eq{opaceq0linearksol}, but with $\gammaa$ being the
Fourier transform of $\Gammaa(x - x^\prime)$, that is, $\gammaa$ defined
by writing
\beq
\label{defgamma}
\Gammaa(x - x^\prime) = \int\frac{d^4k}{(2\pi)^4} e^{-ik\cdot(x - x^\prime)}
\gammaa(k)\,.
\eeq
The fact that $\Gammaa(x - x^\prime)$ is real implies that
\beq
\label{hermiticity1}
\gammaa^\ast(k) = \left.\gammaa(k)\right|_{k\rightarrow -k}\,.
\eeq
Being a scalar, $\gammaa$ is a function of the scalar variables
$\omega$ and $\kappa$ defined in \Eq{omegakappa}, a fact that
we indicate by writing it as $\gammaa(\omega,\kappa)$ when
needed. In particular, \Eq{hermiticity1} actually implies the condition
\beq
\label{hermiticity}
\gammaa^\ast(\omega,\kappa) = \gammaa(-\omega,\kappa)\,.
\eeq
This case includes the original \Eq{opaceq0} (constant $\gammaa$) as
well as \Eq{opaceq1} as special cases, and of course \Eq{correponddencecase1}
is consistent with \Eq{hermiticity}, as it should be.

\subsection{Neutrino background}

An example case in which $\gamma_a$ has the form given in
\Eq{correponddencecase1} is afforded by an electron plasma
with a neutrino background.
The calculation of $\pi_P$ in that case gives\cite{mohanty}
\beq
\pi_P(\omega,\kappa\rightarrow 0) =
\frac{\sqrt{2}G_F\alpha}{3\pi}\left(\frac{\omega^2_{pl,e}}{m^2_e}\right)
(n_{\nu_e} - n_{\bar\nu_e})\kappa\,,
\eeq
where $n_{\nu_e}$ and $n_{\bar\nu_e}$ stand for the number densities of
the electron neutrinos and antineutrinos, and $\omega_{pl,e}$ is the
electron plasma frequency. For simplicity let us consider
the nonrelativistic limit, so that only the electrons (no positrons)
are present, in which case (i.e., \Eqs{defplasmafreq}{defplasmafreqNR})
\beq
\omega^2_{pl,e} = q^2_e\Omega^2_e = \frac{q^2_e n_{e0}}{m_e}\,,
\eeq
where $q_e$ is the electron charge and $n_{e0}$ is the equilibrium
electron number density. On the other hand,
for this case that we are considering, in the framework of the kinetic equation
\beq
\pi_P = \frac{i\kappa q_e \gamma_e\Omega^2_e}{\omega}\,.
\eeq
Therefore, in the framework of the kinetic equation,
the effects of the neutrino background can be parametrized
in terms of a $\gamma_e$ parameter for the electron of the form
\beq
\gamma_e = -i\omega\gamma^{(1)}_e\,,
\eeq
with
\beq
\gamma^{(1)}_e = q_e
\frac{\sqrt{2}G_F\alpha}{3\pi m^2_e}
(n_{\nu_e} - n_{\bar\nu_e})\,.
\eeq

The main lesson here is that the kinetic approach allows
us to parametrize the effects produced by the $\pi_P$ term
in the photon polarization tensor in terms of the parameter
$\gamma_a$. In this framework, $\gammaa$ is a phenomenological parameter
that must be determined by other means, e.g., thermal field theory
in the case of an electron plasma with neutrino background, as we have seen.
Nevertheless, the kinetic approach allows us to study further
the consequences of the presence of the $\pi_P$ term,
such as the effects of external fields\cite{semikoz1},
streaming neutrino background\cite{petropavlova} or
collisional plasmas\cite{rafelski}, among others.
\section{Conclusions and Outlook}
\label{sec:conclusions}

In this work we have proposed a method to study the
propagation of photons in an optically active isotropic medium,
based on the covariant collisionless Boltzmann equation.
As shown in \Section{sec:preliminaries},
the covariant Boltzmann equation can be modified by adding a term that gives
rise to the optical activity effects, in a way that
is consistent with the general requirements of current conservation
and symmetry considerations.
In \Section{sec:linearization}, using the linear
response method, we obtained an expression for the induced current,
expressed in terms of integrals over the
momentum distribution function of the background particles.
There we established contact with the
photon polarization tensor, or equivalently the photon self-energy,
in the TFT language, specifically as used in \Rref{piP},
and the formulas for the components of the polarization tensor
were obtained. The results for the longitudinal and transverse components
of the photon self-energy, $\pi_{L,T}$ respectively, are the familiar ones.
The new result here is the formula for the $P$- and $CP$-breaking component
$\pi_P$ due to the new term we considered
in the Boltzmann equation to describe the effects of optical activity.
In \Section{sec:discussion} we discussed some details of the
results obtained, and considered specifically some particular cases
(e.g., the long-wavelength and the nonrelativistic limit)
that are useful in practical applications and representative of
more general situations.
To emphasize the usefulness of the method,
we pointed out how the $P$- and $CP$-breaking term we included in
the Boltzmann equation can be generalized,
in particular how it applies to a plasma 
with a neutrino gas as a background, and established contact
with calculations of the photon self-energy in such contexts
in the framework of TFT\cite{mohanty}.
The strength and advantages of the method here presented comes from
its semiclassical standpoint, which in many circumstances is more suitable
than the thermal field theory approach
for incorporating other potentially important effects 
such as collisions, external fields, stream backgrounds
and multicomponent plasmas.

\appendix

\section{Notation and conventions}
\label{sec:appendixA}

We use the notation and conventions used in \Rref{piP}, which we briefly
review here for convenience. The momentum of the propagating photon
is denoted by $k^\mu$, and $u^\mu$ is the velocity four-vector
of the medium, already introduced in \Eq{Jcov}.

\subsection{Photon polarization tensor}

In the context of TFT, the photon self-energy, $\pi_{\mu\nu}$
gives rise to a contribution to the effective Lagrangian of the photon
and the corresponding field equation that leads to identify
\beq
\label{jpi}
j_\mu = -\pi_{\mu\nu} A^\nu\,,
\eeq
as the induced current in the presence in the external field,
and whence $\pi_{\mu\nu}$ as the polarization tensor.
As discussed in that reference, the most general form of $\pi_{\mu\nu}$
in an isotropic medium is
\beq
\label{pimunuparam}
\pi_{\mu\nu}(k,u) = \pi_T(\omega,\kappa) R_{\mu\nu}(k,u) +
\pi_L(\omega,\kappa) Q_{\mu\nu}(k,u) + \pi_P(\omega,\kappa) P_{\mu\nu}(k,u)\,,
\eeq
where $\omega$ and $\kappa$ are the scalar variables
\beq
\label{omegakappa}
\omega = k\cdot u\,,\qquad \kappa = (\omega^2 - k^2)^\frac{1}{2}\,,
\eeq
which have the interpretation of the energy and the magnitude of the
momentum of the photon, in the rest frame of the medium.
The tensors $R,Q,P$ are defined as follows. First,
the component of $u^\mu$ transverse to $k^\mu$ is
\beq
\label{tildeu}
\tilde u_\mu = \tilde g_{\mu\nu} u^\nu\,,
\eeq
where
\beq
\label{gtilde}
\tilde g_{\mu\nu} = g_{\mu\nu} - \frac{k_\mu k_\nu}{k^2}\,.
\eeq
Then,
\beqa
\label{RQP}
Q_{\mu\nu} & = & \frac{\tilde u_\mu \tilde u_\nu}{\tilde u^2}\,,\nonumber\\
R_{\mu\nu} & = & \tilde g_{\mu\nu} - Q_{\mu\nu}\,,\nonumber\\
P_{\mu\nu} & = & \frac{i}{\kappa}\epsilon_{\mu\nu\alpha\beta}k^\alpha u^\beta\,.
\eeqa
It is useful to remember that all three tensors are transverse to $k^\mu$,
that is
\beq
\label{RQPk}
k^\mu T_{\mu\nu} = 0 = k^\nu T_{\mu\nu}\,;\quad (T = R,Q,P)\,,
\eeq
and also that $R$ and $P$ are transverse to $u^\mu$ as well,
\beq
\label{RPu}
u^\mu T_{\mu\nu} = 0 = u^\nu T_{\mu\nu}\,;\quad (T = R,P)\,.
\eeq
They satisfy various product relations, among them
\beq
\label{RQPnormalization}
R^{\mu\nu} R_{\mu\nu} = {R^\mu}_\mu = 2\,,\quad
Q^{\mu\nu} Q_{\mu\nu} = {Q^\mu}_\mu = 1\,,\quad
P^{\mu\nu} P_{\mu\nu} = -2\,,
\eeq
and
\beq
\label{RQPproducts}
R^{\mu\lambda}Q_{\lambda\nu} = 0\,,\quad
Q^{\mu\lambda}P_{\lambda\nu} = 0\,,\quad
P^{\mu\lambda}R_{\lambda\nu} = {P^\mu}_\nu\,,\quad
P^{\mu\lambda}P_{\lambda\nu} = {R^\mu}_\nu\,.
\eeq
In the rest frame of the medium, the components of $R_{\mu\nu}$
and $P_{\mu\nu}$ are,
\beqa
R_{00} = R_{0i} = R_{i0} = 0\,,&&\qquad
R_{ij} = \delta_{ij} + \frac{\kappa_i\kappa_j}{\kappa^2}\,,\nonumber\\
P_{00} = P_{0i} = P_{i0} = 0\,,&&\qquad
P_{ij} = \frac{i}{\kappa}\epsilon_{ijk}\kappa^k\,.
\eeqa

\subsection{Dispersion relations}
\label{sec:appendix:dr}

The equation $\partial^\mu F_{\mu\nu} = j_\nu$, in momentum space becomes
\beq
\label{eveqall}
[(k^2 - \pi_T)R_{\mu\nu} + (k^2 - \pi_L)Q_{\mu\nu} -
\pi_P P_{\mu\nu}]A^\nu = 0\,,
\eeq
which determines the dispersion relations and polarization vectors
of the propagating modes. To discuss them we recall the definition of
the transverse vectors $e^\mu_{1,2}$, which in the rest frame of the medium
have components
\beq
e^\mu_{1,2} = (0,\vec e_{1,2})\,,
\eeq
where $\vec e_{1,2}$ are unit vectors with
\beq
\vec e_{1,2}\cdot\hat\kappa = 0\,,\quad \vec e_2 = \hat\kappa\times\vec e_1\,.
\eeq
In covariant form, they satisfy
\beq
\label{RQe12}
R_{\mu\nu} e^\nu_{a} = e_{a\mu}\,,\quad Q_{\mu\nu}e^\nu_{a} = 0\,,
\qquad(a = 1,2)\,,
\eeq
and
\beq
\label{Pe1e2}
e^\mu_2 = -iP^{\mu\nu}e_{1\nu}\,,\quad
e^\mu_1 = iP^{\mu\nu}e_{2\nu}\,.
\eeq
In addition it is useful to introduce
\beq
e^\mu_3 = \frac{\tilde u^\mu}{\sqrt{-\tilde u^2}}\,,
\eeq
which together with $e^\mu_{1,2}$ form a basis in the subspace orthogonal
to $k^\mu$. 

From the fact that $R^{\mu\nu}$ and $P^{\mu\nu}$ acting on $\tilde u_\nu$
give zero, it follows that $A^\mu \sim e^\mu_3$ is a solution of \Eq{eveqall}
provided
\beq
k^2 - \pi_L = 0\,,
\eeq
which is the equation for the dispersion relation $\omega_L(\kappa)$
for the longitudinal mode. Since the presence of the $\gammaa$ term
does not affect $\pi_L$, the dispersion relation for the longitudinal
mode is not affected.

In the absence of the $\pi_P$ term, \Eq{RQe12} implies
that $A^\mu \sim e^\mu_{1,2}$, or any combination of them,
give a solution of \Eq{eveqall} if
\beq
k^2 - \pi_T = 0\,,
\eeq
which gives the dispersion relation $\omega_T(\kappa)$,
The transverse modes, corresponding to the polarization vectors $e^\mu_{1,2}$
are therefore degenerate, with the same dispersion relation $\omega_T(\kappa)$.

As a consequence of the relations in \Eq{Pe1e2}, neither $A^\mu \sim e^\mu_1$
nor $A^\mu \sim e^\mu_2$ are separately solutions of the equation. At this
point it is useful to introduce the circular polarization vectors
\beq
e^{(\pm)\mu} = \frac{1}{\sqrt{2}}(e^\mu_1 \pm ie^\mu_2)\,,
\eeq
which satisfy
\beq
P^{\mu\nu}e^{(\lambda)}_\nu = \lambda e^{(\lambda)\mu}\,,
\quad \lambda = \pm\,,
\eeq
in addition to identities analogous to \Eq{RQe12}. It then follows that
$A^\mu \sim e^{(\pm)\mu}$ are each a solution of the equation, with
the corresponding dispersion relation being the solution of
\beq
\label{dreq1}
k^2 - (\pi_T + \lambda\pi_P) = 0\,.
\eeq

\subsection{Dielectric tensor}
An equivalent way to express the presence of the $\pi_P$ term
in the photon self-energy is in terms of the
components of the dielectric tensor, which are given by\cite{piepsilonrelation}
\beq
\label{piepsilonrelations}
1 - \epsilon_t = \pi_T/\omega^2,\quad
1 - \epsilon_\ell = \pi_L/k^2,\quad
\epsilon_p = \pi_P/\omega^2\,.
\eeq
The interpretation is that, in the rest frame of the medium, the
induced current vector is given by
\beq
\vec j = i\omega[(1 - \epsilon_\ell)\vec E_\ell +
  (1 - \epsilon_t)\vec E_t - i\epsilon_p\hat\kappa\times\vec E]\,,
\eeq
where we are writing, in that frame, 
\beq
k^\mu = (\omega,\vec\kappa)\,,
\eeq
while $\vec E_\ell$ and $\vec E_t$ denote the components
of the electric field parallel and transverse to $\vec\kappa$, respectively.
The $\epsilon_p$ term breaks the degeneracy between the two transverse
polarization states of the propagating photon.


\end{document}